\documentclass[%
 preprint,
 amsmath,amssymb,
 aps,
]{revtex4-2}

\usepackage{graphicx}    
\usepackage{bm}          
\usepackage{dcolumn}     
\usepackage{booktabs}    
\usepackage{subcaption}  
\usepackage{xcolor}      
\usepackage{hyperref}    
\usepackage{physics}     
\usepackage{siunitx}     

\hypersetup{
 colorlinks=true,
 linkcolor=blue,
 citecolor=blue,
 urlcolor=blue
}

\usepackage[normalem]{ulem}

\begin{document}

\title{Coalescence of Polymer Droplets Moving on a Surface with Stiffness Gradient}


\author{Divyansh Tripathi$^{1}$}
\email{divyanshtripathi@bhu.ac.in}

\author{Vimal Kishore$^{1}$} 
\email{vimalk@bhu.ac.in}

\author{Panagiotis E. Theodorakis$^{2}$}

\author{Swarn Lata Singh$^{3}$} 
\email{swarn@bhu.ac.in}

 \affiliation{$^{1}$ \quad Department of Physics, Banaras Hindu University, Varanasi 221005, India \\
  $^{2}$ \quad Institute of Physics, Polish Academy of Sciences, Al. Lotników 32/46, 02-668 Warsaw, Poland \\
 $^{3}$ \quad Department of Physics, Mahila Mahavidyalaya, Banaras Hindu University, Varanasi 221005, India }

\date{\today}


\begin{abstract}
 Coalescence of sessile droplets has been in the focus of current and previous research, due to its relevance for various biological processes and industrial applications. However, the coalescence of moving droplets on substrates with varying properties, such as gradient substrates, has received less attention. Hence, the main focus here is on the coalescence of droplets that are moving in the same direction on a soft surface; the motion of the droplets is caused by a gradient in the surface stiffness. As reference, stationary coalescence of the same droplets is also studied on the corresponding uniform surfaces for different stiffness values. To describe the coalescence phenomenon on a surface with stiffness gradient, a relevant range of velocity ratios of the leading and the trailing droplet was considered to elucidate the effect of this parameter on coalescence. Moreover, to analyze the dynamics of the process, the temporal growth of the bridge height $(h)$ was investigated, which follows a power law $(h \sim t^{\alpha})$, before eventually attaining a constant value. The obtained values of $\alpha$ show a transition from a higher to a lower value as a function of time, pointing to the presence of two distinct power-law growth regimes, where the transition signifies the crossover from the capillarity-dominated regime to the viscoelasticty-dominated regime of coalescence. In addition, varying attractive strengths for droplet--droplet and intra-droplet interactions were considered. The results indicate that both the dynamics and the degree of the coalescence strongly depend on these interaction parameters. Thus, we anticipate that our results will shed more light on the durotaxis-driven coalescence of polymeric droplets for various relevant system parameters, which will have practical implications for applications ranging from microfluidics to ink-jet printing, where substrate properties may vary. In addition, results may add to the fundamental understanding of the interactions among multicellular aggregates moving on biological surfaces.
\end{abstract}

\maketitle
\setlength{\abovedisplayskip}{2pt}
\setlength{\belowdisplayskip}{2pt}

\section{Introduction}{\label{intro}}
Droplet interactions are of fundamental importance in a wide range of industrial applications.
The texture and the stability of emulsions, which make the very basis of various food, cosmetics, and pharmaceutical products, greatly depends on the interaction among constituent droplets~\cite{ Li2023,Xiao2025,TuckerMoldenaers2002}. Further examples include but are not limited to applications in microfluidics, spray cooling, coating, printing and oil--water separation \cite{Kamp2016,Sarig2016,Pradhan2021,Hafskjold1994}. Interacting droplets have also gained attention as a model system to study the interaction between cells and cellular aggregates \cite{Pizarro2022,Meredith2020}. Such studies are also helpful in gaining insights into biological processes such as the interaction of immune cells with pathogens, and the formation of tissue and growth of bacterial colonies~\cite{Pnisch2017,Foty2005,Oriola2022,Sart2014,Oakes2009,Jain2019}. In such cases, fundamentally understanding the effect of relevant system parameters that dictate the interplay between the droplets could lead to better droplet control in applications ranging from microfluidics ~\cite{Li2020}, to biomedical~\cite{Whitesides2001} and chemical applications~\cite{Geng2017}. In those, the various parameters influencing the droplet behavior include the fluid properties of the droplets, their mutual interactions, geometries (whether the droplet is sessile, or pendant), their approach velocities, and  the presence of external forces~\cite{Zhao2025, Jaccorev}.  

 The capillarity driven merging of two or more droplets to form a single droplet is one of the most well studied phenomena in the literature~\cite{EGGERS1999,Eggers2025,Yue2024,MenchacaRocha2001,Ryu2023,Kavehpour2015,Aarts2005,Arumugam2024}. In a rather simple scenario, namely the coalescence of freely suspended, miscible droplets of a Newtonian fluid, the dynamics of the coalescence process can be described through the rate of the radius growth of the liquid bridge, which forms between the coalescing droplets. In particular, it has been shown that the temporal growth of the bridge radius follows a power-law scaling with time. However, the exponent of the scaling is not universal, since it depends on various system parameters. For example, for freely suspended Newtonian droplets, the radius of the bridge grows linearly with time $(\propto t)$ in the viscous regime, where the resistance to coalescence is predominantly attributed to viscous forces, while the role of the inertia can be neglected \cite{Hopper1984,Hopper1990}. In contrast, in the inertia dominated regime, the bridge radius growth is proportional to $t^{0.5}$
\cite{EGGERS1999,DUCHEMIN2003}. A third regime has also been reported in the literature, namely the inertially limited viscous (ILV) regime, which occurs when both viscosity and inertia are comparable. Some studies have reported ILV  as the initial regime for all coalescence events \cite{Paulsen2012}. A transition from the ILV behavior to either the viscous 
 or the inertial regimes depends on the fluid properties of the droplet and its size ~\cite{Chandra2025,Paulsen2012,Paulsen2011,Paulsen2013,Joshi2025}.

Only a few studies have thus far concerned interacting, non-Newtonian droplets in the literature. In this case, various factors are supposed to affect the dynamics of coalescence, such as the elastic properties of the droplets and the intrinsic time scale for relaxation \cite{Chandra2025}. Here, examples of  non-Newtonian fluids range from polymeric solutions to liquid-crystal and biological soft matter, where structural properties at microscopic level substantially vary for each sub-class. As a result, a unique response is expected for each of these materials, which renders difficult the formulation of a unified framework that could describe the coalescence in the case of non-Newtonian droplets. For example, polymeric droplets, which are relevant for technological applications and, also, often serve as simplified model systems for investigating biofluids, are among the most well studied in the case of complex fluids. Still, various results concerning the coalesceence of polymeric droplets are contradictory in the literature, which calls for more research on the subject. On the one hand, for example, Dekker {\it et. al,} \cite{Dekker2022} have reported that the neck radius for coalescing (free hanging) polymer droplets follows the same temporal evolution as that of Newtonian droplets of very low viscosity, {\it i.e,} $\beta = 0.5$, where $\beta$ is an exponent expressing  the temporal evolution of the neck radius. In this case, the presence of polymer chains was shown to affect the spatial features of coalescence. On the other hand, Verma {\it et.al} \cite{Varma2020} reported a value of $\beta = 0.36$ for the evolution of the neck radius of a sessile -- pendant droplet coalescence. Moreover, the value of $\beta$ decreased even further for polymer concentrations roughly $20$ times more than the critical concentration.

The coalescence of sessile polymeric droplets is relevant for many industrial applications such as microfluidics~\cite{Hassan2021}, spray coating~\cite{Lohse2022}, and $3D$ printing~\cite{Klestova2019}. For coalescing droplets placed on a solid surface, the substrate introduces additional viscous stress. This viscous stress which depends on the wettability of the substrate and, can significantly affect the interactions. The coalescence dynamics can be quantified in terms of the temporal growth of the bridge width, $r$, and the bridge height, $h$. Similarl to the previous case, these two parameters grow in time as power laws which are expressed as follows: $r(t) \propto t^{\beta}$ and $h(t) \propto t^{\alpha}$. Dekker {\it et. al,} \cite{Dekker2022} has reported the same time evolution for bridge height as that of sessile inviscid Newtonian droplets ($\alpha = 2/3$). A departure from the Newtonian regime was observed only at very high polymer concentrations. Verma {\it et. al,} {\cite{Varma2021} studied the coalescence of polymeric fluid droplets on a partially wettable substrate, and a transition from $\alpha = 2/3$ to $\alpha = 1/2$ was observed with increasing polymer concentration. Arbabi {\it et. al,} ~\cite{Arbabi2023} studied the coalescence of sessile polymeric droplets on a substrate with two different contact angles; $78^{\circ}$ and $118^{\circ}$. The value of $\alpha$ was found to be greater at $\theta = 78^{o}$ than that in the case of $\theta = 118^{o}$, when all other parameters were kept the same. It was also observed that $\alpha$ decreases with increasing viscosity. For all the viscosities considered for both contact angles, the values of $\beta$ were found to be less than $0.5$ \cite{Arbabi2023}. Rostami {\it et. al,} \cite{Rostami2025} have reported values of $\alpha$ and $\beta$ for coalescing 
sessile viscoelastic droplets located on a hydrophobic surface, for varying droplet elasticity. The value of $\alpha$ first decreases with increasing elasticity and reaches a minimum for the regime when both elasticity and viscosity play a role. When the elasticity increases further, the value of $\alpha$ increases again and reaches a plateau. All the reported values of $\alpha$, again, were $ < 0.5$. Kaneelil {\it et. al,} \cite{Kaneelil2026} have reported  $\alpha \approx 1$ for two sessile viscoelastic droplets in the small contact-angle regime. This value was unaffected by varying the polymer concentration. Moreover, $\alpha$ was found to decrease gradually from an initial value of $\alpha = 2/3$ as the system moved from the viscosity dominant to the elasticity dominant regime.While most of these studies concern droplets coalescence on rigid substrates, only some of them have considered soft substrates. For example, Roy {\it et.al,} \cite{Roy2020} studied the coalescence of water droplets on a soft surface and has concluded that droplets exhibit reluctance to coalescence on these soft substrates. Sokuler {\it et. al,} \cite{Sokuler2009} also studied the merging of water droplets on a soft surface with increasing softness. They have reported similar results, namely, the relaxation time for drop shape after merging increases with increasing softness. They also observed that coalescence was prevented on surfaces, when the softness was higher than a threshold.

Viscoelastic droplets located on a soft surface is a common occurrence both in nature and technology. For soft surfaces the resting droplet may create deformations underneath, which, in turn, affects the coalescence. Furthermore, these deformations could be anisotropic and may lead to a directional motion of the droplet for a soft substrate with stiffness gradient \cite{Kajouri2024,Theodorakis2017}. Such processes are also relevant in the context of cellular clusters. For example, they can migrate in a directional manner following the gradient in the mechanical properties of the substrate. Moreover, Understanding the behavior of multiple droplets moving on such deformable substrates may enhance our understanding about the fusion of cellular aggregates, which is an important step in morphogenesis, cancer metastasis and cell-based therapies \cite{Palmiero2023,Kim2013}. In addition, such a system setup may also be relevant for explaining the growth of microbial communities, on surfaces, through aggregate fusion under various conditions \cite{Porter2025,Pnisch2017,Welker2018}. 

In this regard, we study the mutual interaction between two polymeric droplets that perform durotaxis motion on a gradient substrate \cite{Theodorakis2026}. Since the motion of the droplets takes place along the stiffness gradient, the interacting droplets essentially move in the same direction leading to a coupled interplay of droplet motion and coalescence on a substrate of varying wettability (stiffness) (Fig.\ref{fig:scheme}). We anticipate that our study not only enhances  understanding the fusion and growth of multicellular aggregates on a surface, but is also relevant for microfluidics, and biomedical applications~\cite{Zhu2025,Alistar2017,SungKwonCho2003,Knoche1994,Pnisch2017}. The manuscript is organized as following; in the section~\ref{system}, we discuss the system that is studied, in details and we describe the the computational method used to carry out the investigation. In next, we discuss the results in the section~\ref{result} which is followed section~\ref{summary} where we conclude the manuscript.

\section{System, Model and Methods}{\label{system}}
\begin{figure}[h!]
    \centering
    \includegraphics[width=0.5\linewidth]{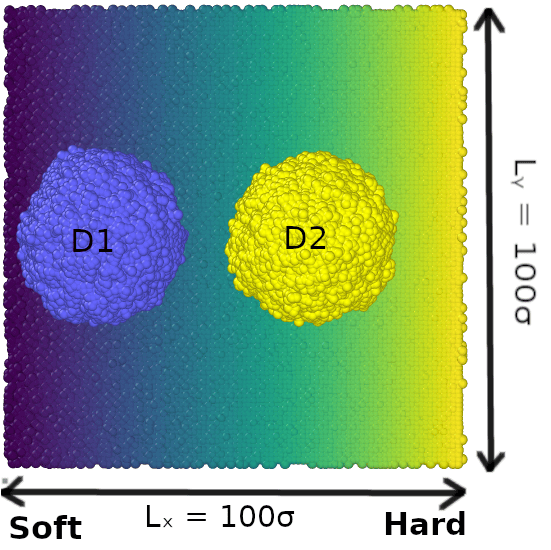}
    \caption{ System setup consisting of two polymer droplets marked as $D_{1}$ and $D_{2}$, which are placed on a soft surface with stiffness gradient in positive $x-$direction as indicated by the color code and the arrow. In the case of stationary coalescence, the surface stiffness was kept uniform.}
    \label{fig:scheme}
\end{figure}
Our
model is built on previous studies of the durotactic droplet motion of a polymer droplet on a
surface with stiffness gradient,~\cite{Theodorakis2017}. Here, our system consists of two polymer droplets placed on the substrate,
while in the case of stationary coalescence the substrate stiffness was uniform.  To simulate the coalescence between two droplets, which move in the same direction, a stiffness gradient was applied on the surface by using a harmonic potential as has been done previously~\cite{Theodorakis2017}. This harmonic interaction potential reads

\begin{equation}
U_{harmonic(r)} = -\frac{1}{2} K r^2 ,
\end{equation}
where $r = 0$, when the bead is at its equilibrium position, and $K$ is the spring constant, which is gradually changing to implement the stiffness gradient of the substrate. In particular,
small values of $K$ result in softer substrates (large thermal fluctuations of the beads) with the softness of the surface decreasing} 
with increasing values of $K$. To create the gradient, the stiffness is increased by $\Delta K$ at every $\Delta L = 4\sigma$ ($\sigma$ is the unit of length), starting from an initial value of $K_0$. This increment takes place along the positive $x-$axis, starting from $x = 0$ to $x = 100\sigma$ (a top view of the system is presented in Fig.\ref{fig:scheme}){ \cite{Theodorakis2017}}. The linear dimensions of the substrate in both the $x$ and the $y$ directions are set to $L_{x} = L_{y} = 100\sigma$.

The droplets are made up of polymer chains, each containing $10$ monomers. Consecutive monomers of each polymer chain are bonded by using the finite extensible nonlinear elastic (FENE) potential, which is mathematically expressed as follows:

\begin{equation}
U_{\mathrm{FENE}}(r)
=
-\frac{1}{2} k R_0^2
\ln\!\left[1 - \left(\frac{r}{R_0}\right)^2\right],
\end{equation}
where $r$ is the distance between two bonded monomers,  $k=30.0$ $\varepsilon/\sigma^{2}$ is the elastic constant, and $R_{0}=1.5\sigma$ the maximum bond extension. All non-bonded interactions between beads belonging to both the droplets and the substrate were modeled using the truncated and shifted Lennard-Jones (LJ) potential
\begin{equation}
U_{ij}(r_{ij}) = 4 \varepsilon_{ij} \left[ \left( \frac{\sigma_{ij}}{r_{ij}} \right)^{12} - \left( \frac{\sigma_{ij}}{r_{ij}} \right)^{6} \right],
\end{equation}

where, $i$ and $j$ stand for the type of monomer. $d_{1}$ denotes monomers making up the droplet $D_{1}$, $d_{2}$  indicates the monomers belonging to droplet $D_{2}$, while $s$ stands for the monomers constituting the surface. $r_{ij}$ is the distance between every interacting pair of beads  within a
cut-off distance, which is set to $r_{c} = 2.5\sigma$ for the droplet--droplet and the droplet--surface interactions, while $r_{c} = 2^{1/6} \sigma$ for the LJ interactions between the  substrate beads.  In the present study, we kept $\varepsilon_{ss} = \varepsilon_{d_{1}d_{1}} = \varepsilon$ fixed ($\varepsilon$ is the energy unit),  while $\varepsilon_{d_{1}s}$, $\varepsilon_{d_{2}s}$, $\varepsilon_{d_{2}d_{2}}$,  and $\varepsilon_{d_{1}d_{2}}$  were varied to investigate the various scenarios. 

The molecular dynamics (MD) simulations were conducted using LAMMPS package\cite{LAMMPS}. The Langevin thermostat, which was used here, includes random and dissipative forces to the Newton's equation of motion for each bead, namely:
\begin{equation}
m\frac{d^2{\bf r}_{i}}{dt^2} = -\nabla U_{i} -\gamma\frac{d{\bf r}_{i}}{dt} + \Gamma_{i}(t)
\end{equation}
where $m =  \rm m$ is the mass of each monomer (all the monomers are of the same mass considering the same mass unit, $\rm m$, for all cases), $\nabla U_{i}$ is the force acting on the $i^{th}$ monomer due to the presence of all other beads within the cut-off distance, and $\gamma$ is the damping coefficient set to $0.1 {\tau}^{-1}$,  where $\tau = (\frac{\rm m\sigma^{2}}{\varepsilon})^{1/2}$ 
is the time unit. The Langevin equation is integrated for each bead using the velocity-Verlet algorithm with time step
$\delta t = 0.005 \tau$. $\Gamma$ is the random force on each bead representing the thermal 
fluctuations, and is related to $\gamma$ via the fluctuations--dissipation theorem, namely
\begin{equation}
\langle \Gamma_{i}(t)\Gamma_{j}(t^{'})\rangle = 6k_{B} T \gamma \delta_{ij} \delta(t - t^{'}) .
\end{equation}

The dimensions of the simulation box are chosen large enough to avoid interactions with mirror images or the system,  due to the presence of periodic
boundary conditions in all three dimensions. Both droplets are equilibrated on the surface,
keeping their mutual interaction off, for long enough such that the average shape of the droplets 
does not change with time. The droplets are placed along the stiffness gradient (positive $x-$axis)
maintaining the distance between the last bead of the first droplet and the first bead of the last droplet to a value higher than $2.5\sigma$ (cutoff distance) after the equilibration. Once equilibrium was reached, the mutual interaction between the droplets was turned on. As a result, both droplets started moving due to the stiffness gradient, and subsequently are able to get close enough to start to interact or coalesce. 

\section{Results}{\label{result}}

In this section, we present an analysis of the outcomes of the simulation runs for coalescence under varying system conditions. Temperature, FENE potential parameters, and the droplet size were kept fixed throughout the study. Temperature is set to be $T = 1.0\,\varepsilon/k_{\mathrm{B}}$ , the FENE potential parameters are fixed at $k=30.0$ $\varepsilon/\sigma^{2}$, and $R_{0}=1.5\sigma$. The coalescing droplets are both made up of $1800$ polymers; each polymer contains $10$ monomers. All the distances, in the following, are given in the units of $\sigma$ and the potential parameters for well depth are in units of $\varepsilon$. 
We studied coalescence of stationary as well as moving droplets on
a surface. For stationary coalescence, three different values of surface stiffness were considered. For the coalescence of moving droplets, the droplets were placed on a stiffness gradient surface. The two droplets, then perform durotactic motion along the stiffness gradient and, somewhere on the way, these droplets meet and merge. This amounts to coalescence of two droplets moving in the same direction with trailing droplet moving with a higher speed than that of the leading droplet. A range of velocities ratio of trailing and leading droplets have been deployed to decode the effect of droplet's motion on the coalescence.

Also, the strength of adhesion and inter-droplet cohesion was varied to gain an insight into the role of various intermolecular interactions on the process. In Fig.~$1$, we present the scheme of simulated system. There are two droplets, $D_{1}$ and $D_{2}$, placed on a soft surface. For stationary coalescence, the substrate's spring constant remains fixed throughout the surface. The stiffness of the surface increases with increasing $K$ value, whereas the contact angle made by the droplet on surface decreases. In Table$1$, we have listed the values of $\theta_{Y}$ for each combination of $K$ and $\varepsilon$.

\subsection{Coalescence of stationary droplets on a soft surface}{\label{resulta}}

\begin{table}[h!]
\centering
\renewcommand{\arraystretch}{0.7}
\caption{Dependence of contact angle on droplet--substrate interaction strength and substrate stiffness.}
\label{tab:contact_angle}
\begin{tabular}{|c|c|c|}
\hline
 \textbf{Interaction Strength ($\varepsilon_{ds}$)} &  \textbf{Substrate Stiffness (k = $\varepsilon/\sigma^2$)} &  \textbf{Contact Angle ($~\sim \theta^\circ$)} \\
\hline
 & 20  &  $133^\circ$ \\
0.4 & 120 &  $122^\circ$ \\
 & 240 &  $120^\circ$ \\
\hline
 & 20  & $122^\circ$ \\
 0.5 & 120 &  $110^\circ$ \\
 & 240 & $106^\circ$ \\
\hline
 & 20  & $111^\circ$  \\
0.6 & 120 &  $96^\circ$ \\
 & 240 &  $94^\circ$ \\
\hline
 & 20 & $97^\circ$ \\
 0.7 & 120 & $82^\circ$ \\
  & 240 & $80^\circ$ \\
  \hline
\end{tabular}
\end{table}
We first discuss the coalescence of $D_{1}$ and $D_{2}$ on uniform substrate's stiffness for three different values of spring constant $K$ , namely $K = 20$, $K = 120$ and $K = 240$. For all three cases, we kept $\varepsilon_{d_{1}s} = 0.7$ and $\varepsilon_{d_{2}s} = 0.40$ in the unit of $\varepsilon$. The values of contact angles for each combination of $K, \varepsilon$ are given in Table~\ref{tab:contact_angle}, where it can be seen that the contact angle $(\theta_{Y})$ decreases with increasing $K$ and $\varepsilon_{ds}$. The values of $\theta_{Y}$ are approximated by fitting spherical caps to the droplets \cite{Heine2004,Allen2003,Lubarda2011,iwamoto2013,Theodorakis2015}.
\begin{figure}[ht]
    \centering

    \begin{minipage}[t]{0.48\linewidth}
        \centering
        {\large \boldmath{$K = 20$}}\\[4pt]

        \includegraphics[width=\linewidth]{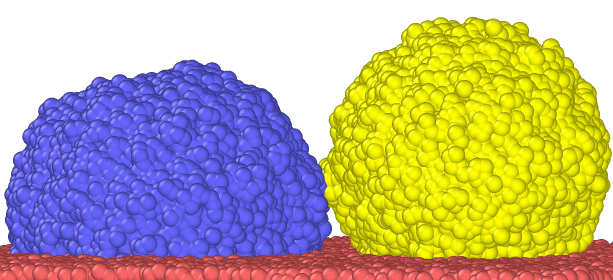}\\
        {\small $t = 2100$ ts}\\

        \includegraphics[width=\linewidth]{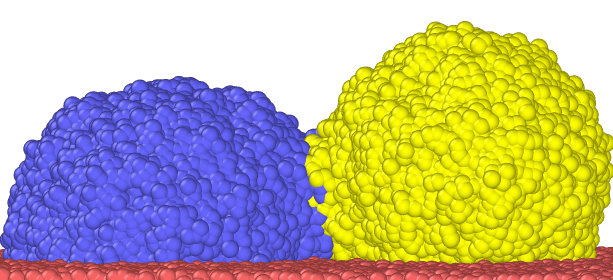}\\
        {\small $t = 1.7 \times 10^4$ ts}\\

        \includegraphics[width=\linewidth]{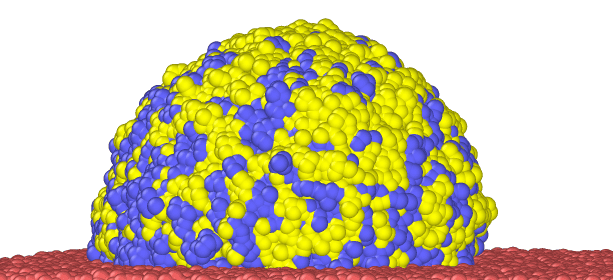}\\
        {\small $t = 29.661\times10^5$ ts}
    \end{minipage}%
    \hfill
    \begin{minipage}[t]{0.48\linewidth}
        \centering
        {\large \boldmath{$K = 240$}}\\[4pt]

        \includegraphics[width=\linewidth]{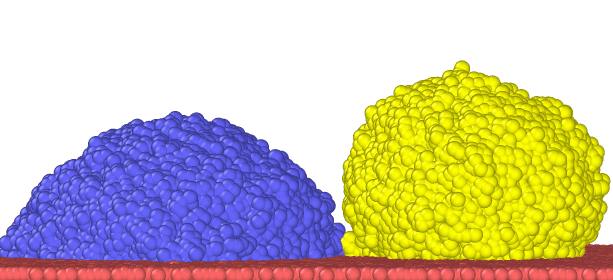}\\
        {\small $t = 2100$ ts}\\

        \includegraphics[width=\linewidth]{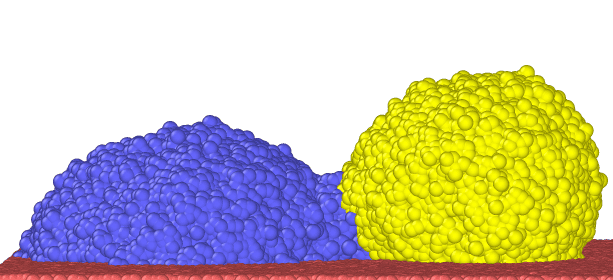}\\
        {\small $t = 4.66\times10^4$ ts}\\

        \includegraphics[width=\linewidth]{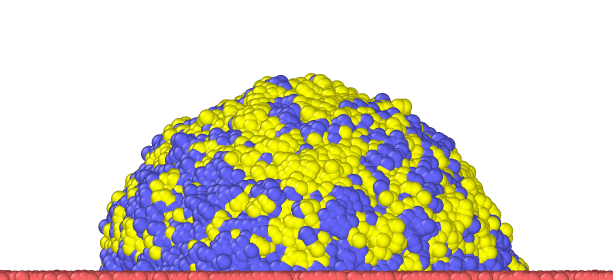}\\
        {\small $t = 51.03\times10^5$ ts}
    \end{minipage}

    \caption{Time evolution of droplet coalescence on a uniform substrate for different stiffness values. For left column, $K=20$, and the substrate is softer than the one on the right column with $K=240$.  In the case of softer substrates the droplets are more spherical and the coalescence starts at the equator of the droplets, whereas for stiffer ones the droplets are more flat and the coalescence starts at the surface. Time is taken in units of timesteps (ts) and the origin is taken when the coalescence starts. The rest of the parameters are same as mentioned in Section~\ref{resulta}.}
    \label{fig:time_evolution}
\end{figure}
In the beginning of the simulation, centers of $D_{1}$ and $D_{2}$ are placed 
such that the last beads of $D_{1}$ and first beads of $D_{2}$ are $\approx 10\sigma$ away. The droplets are equilibrated keeping their mutual interaction off {\it i.e,} $\varepsilon_{d_{1}d_{2}} = 0$. Once equilibrated, the mutual interaction between the droplets is switched on and set to $\varepsilon_{d_{1}d_{2}} = 1.0$ for all the three $K$ values. Overall, $D_{1}$ is attracted more to the surface than $D_{2}$ and hence acquires a flatter shape with smaller $\theta_{Y}$. 

\begin{figure}
    \centering
    \includegraphics[width=0.7\linewidth]{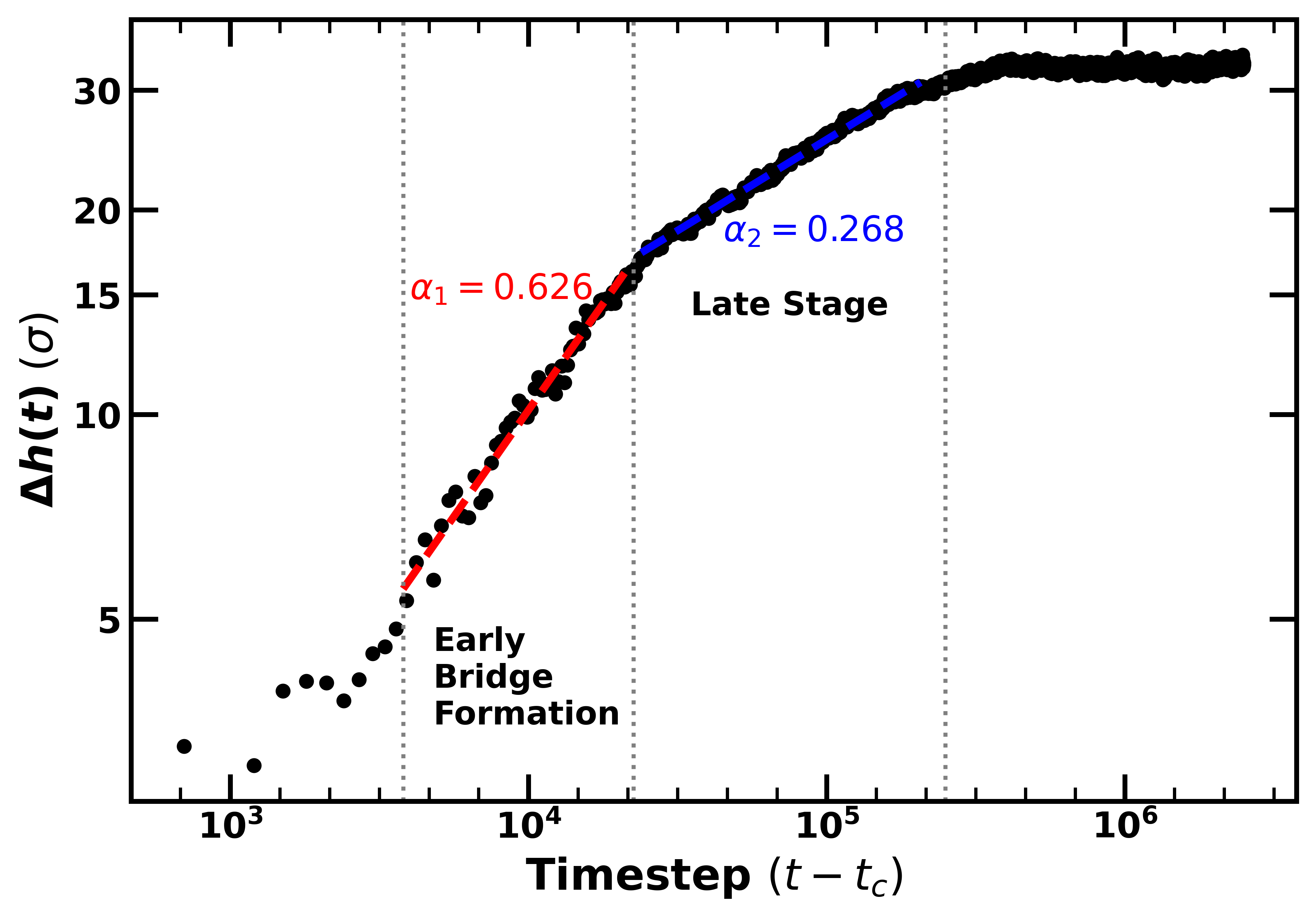}
    \caption{Temporal evolution of bridge height for stationary coalescence on a surface with spring constant $K=20$. The extracted power-law exponent has two values, namely $\alpha_{1} = 0.626$ and $\alpha_{2} = 0.268$, indicating a shift from a faster to a slower bridge growth as the coalescence progresses. The rest of the parameters are the same as mentioned in Sec.~\ref{resulta}.}
    \label{fig:scalk20}
\end{figure}

For $K= 20$, the coalescence starts at the equator of the droplets as the droplets are more spherical and the beads at equators are the closest, namely within $2.5\sigma$. The 
$D_{1}-D_{2}$ interface, or the bridge forms at the equator and then  grows vertically with time
( Fig.~\ref{fig:time_evolution}). This temporal evolution of the bridge height, shown
in Fig.~\ref{fig:scalk20}, is one of the measures to quantify the coalescence dynamics. The bridge height follows a power-law growth in time before reaching a stable value. The power-law growth region consists of two regimes, where the bridge growth is faster in the early stages with a higher value of power-law exponent $(\alpha_{1})$. The dynamics becomes considerably slower at a later stage as the system evolves over time and the corresponding value of the exponent, $(\alpha_{2})$, decreases (Figure~\ref{fig:scalk20}). Such a cross-over in dynamics is anticipated for viscoelastic droplets, since, in the beginning, the polymer chains follow the deformations caused by capillarity and contribute to the bridge growth.   However, once the chains start to relax, the viscoelastic resistance slows down the bridge growth
\cite{Arbabi2023, Rostami2025,Katre2026}.

For $K = 20$, $\alpha_{1} = 0.6260$ and $\alpha_{2} = 0.2679$.  The values of $\alpha_{1}$ decrease with increasing $K$, whereas, $\alpha_{2}$ increases (values are listed in Table~\ref{tab:table_stiffness}). The variation of $\alpha_{1}$ with $K(\theta_{Y})$ can be understood in terms of competition between the droplet--droplet and droplet--surface attractions. For $K =20$, the droplet--surface attraction is weaker and the coalescence starts 
a few molecular diameters away from the substrate. Thus, the competing attraction from the surface, which essentially pulls the polymers forming bridge towards itself, is less effective. 
As $K$ increases, the surface--droplet attraction increases for both droplets, thus the droplets spreading more. The coalescence starts on the surface as well. This stronger attraction competes with the bridge growth in the vertical direction,  thus reducing the values of $\alpha_{1}$. In the second regime of the bridge growth, this trend is reversed, {\it i.e,} the value of $\alpha_{2}$ is the smallest for $K =20$ and increases with $K$. In this regime, 
the viscoelastic relaxation of the polymers is at work, trying to bring the stretched polymers back to equilibrium. In such a scenario, the droplet--surface attraction counters this viscoelastic resistance, and, thus, further favors the bridge growth. This implies that a higher droplet--surface attraction would lead to faster bridge growth in the second regime.
\begin{table}[b!]
\centering
\caption{
Scaling exponents ($\alpha_1$,$\alpha_2$) describing the time evolution of the bridge height, $h(t) \sim t^{\alpha}$, for varying substrate stiffness $k$ at fixed droplet–surface interaction parameters $(\varepsilon_{d_1s}, \varepsilon_{d_2s}) = (0.7,\,0.4)$.The early time exponent $\alpha_1$ is associated with the early bridge formation regime, while the exponent $\alpha_2$ corresponds to the late vertical growth of the interfacial contact region.Uncertainties denote the standard error of the mean.
}
\label{tab:table_stiffness}

\setlength{\tabcolsep}{10pt}
\renewcommand{\arraystretch}{1.2}

\begin{tabular}{|c|c|c|}
\hline
$k$
& Bridge formation (early) $\alpha_1$
& Interfacial growth (late) $\alpha_2$ \\
\hline
20  & $0.6260 \pm 0.0066$ & $0.2679 \pm 0.0006$ \\
120 & $0.4412 \pm 0.0041$ & $0.2899 \pm 0.0007$ \\
240 & $0.4266 \pm 0.0020$ & $0.3089 \pm 0.0008$ \\
\hline
\end{tabular}

\end{table}

As mentioned in Section~\ref{intro}, a few studies have thus far been conducted on the coalescence of sessile polymer droplets, including polymer solution droplets. Moreover, the reported values of $\alpha$ vary from one study to another, although all these studies report $\alpha < 0.5$ in the case of coalescence on partially wettable surfaces. $\alpha > 0.5$ was only reported for the cases where either the polymer concentration, or the contact angle is very small \cite{Dekker2022, Arbabi2023,Rostami2025,Kaneelil2026}. Results presented in our study are comparable to the values reported in the literature, especially the values of $\alpha_{2}$. It is to be emphasized here that, for viscoelastic fluids, two different regimes should exist as a signature of the crossover from the capillarity-dominated to the viscoelasticity-dominated regime \cite{Rostami2025}.

\subsection{Coalescence of moving droplets}{\label{resultb}}
With the understanding  unfolded in the last section, we now turn our attention to the coalescence of two droplets that perform durotactic motion in the same direction on a gradient surface (Fig.~\ref{fig:scheme}). To create the gradient surface, the initial value of the stiffness is set to be $K = 20$ at $x=0$, and is increased by $\Delta K = 10.666$ every $4\sigma$ in the positive $x-$direction. In response to the gradient, the droplets move from the softer to the stiffer edge of the substrate. The speed of the droplets depends on the strength of the droplet--substrate interaction, \textit{i.e.}, $\varepsilon_{d_{1}s}$ and $\varepsilon_{d_{2}s}$. Overall, higher interactions yield higher speeds. The rest of the system parameters are set to  $\varepsilon_{d_{1}s} = 0.7$,  $\varepsilon_{d_{2}s} = 0.4$, and $\varepsilon_{d_{1}d_{1}} = \varepsilon_{d_{2}d_{2}} = \varepsilon_{d_{1}d_{2}} = 1.0$. $\varepsilon_{d_{1}s} > \varepsilon_{d_{2}s}$ ensures that the trailing droplet meets the leading one before reaching the 
end of the substrate.

\begin{figure}[h]
    \centering
    \includegraphics[width=0.65\linewidth]{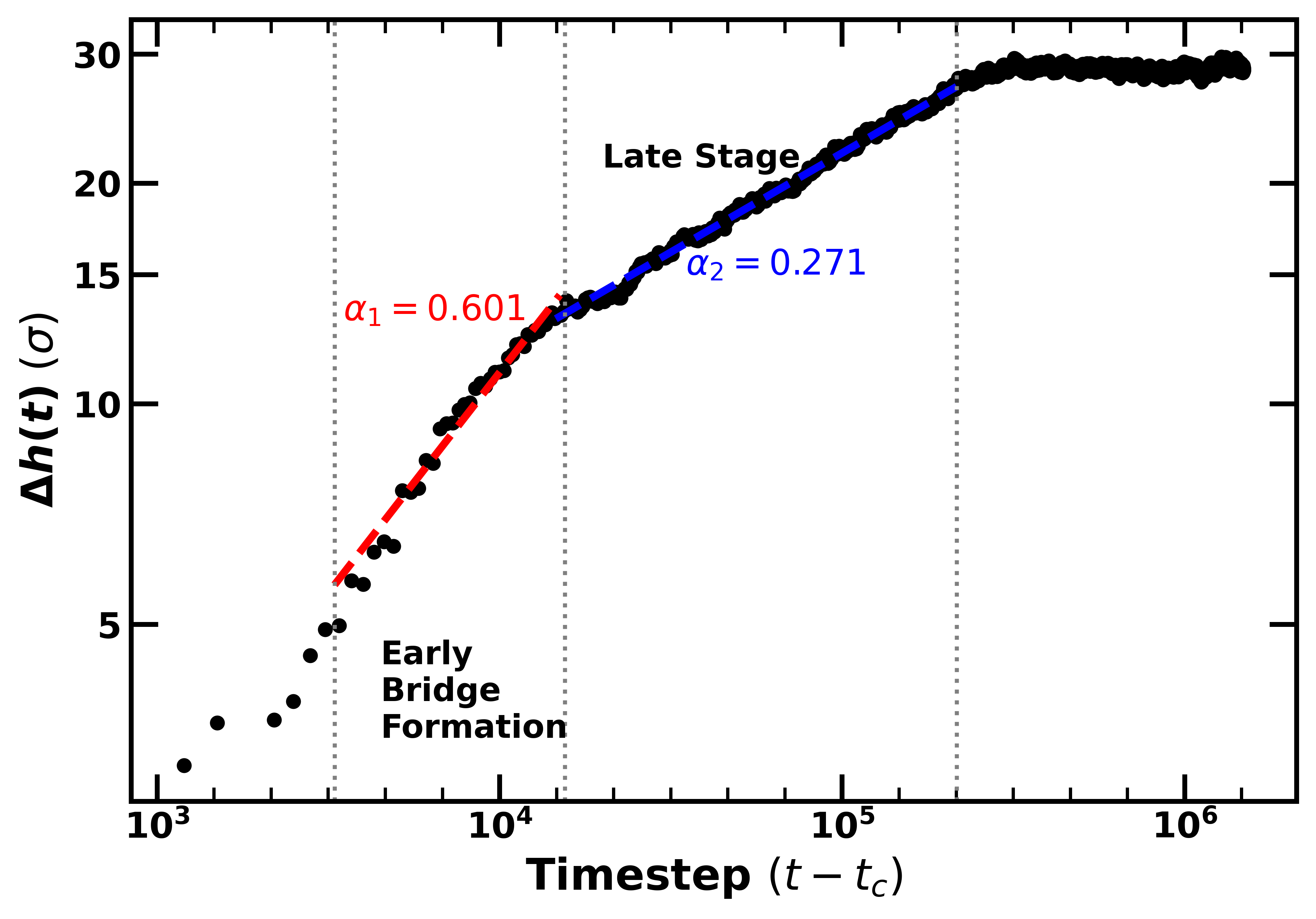}
    \caption{Temporal evolution of bridge height for the coalescence of two droplets 
    moving on a gradient surface with $\varepsilon_{d_{1}s} = 0.70$ and 
    $\varepsilon_{d_{2}s} = 0.40$. The rest of the interaction parameters are same as mentioned in the Sec.~\ref{resultb}. The bridge height, again, has two regimes of power law growth 
    before attaining a stable value. }
    \label{fig:Fig4}
\end{figure}
\begin{table}[htbp!]
\centering
\caption{
Scaling exponents $\alpha$ describing the time evolution of the bridge height, $h(t) \sim t^{\alpha}$, for varying substrate interaction pairs $(\varepsilon_{d_1s}, \varepsilon_{d_2s})$. 
}
\label{tab:table1}
\setlength{\tabcolsep}{9pt}
\renewcommand{\arraystretch}{1.05}
\begin{tabular}{|c|c|c|}
\hline
$(\varepsilon_{1s}, \varepsilon_{2s})$
& Bridge formation (early) $\alpha_1$
& Interfacial growth (late) $\alpha_2$ \\
\hline
$(0.7,\,0.4)$ & $0.6013 \pm 0.0108$ & $0.2707\pm 0.0004$ \\
$(0.6,\,0.4)$ & $0.6342 \pm 0.0097$ & $0.2653 \pm 0.0019$ \\
$(0.5,\,0.4)$ & $0.6509 \pm 0.0079$ & $0.2454 \pm 0.0015$ \\
$(0.4,\,0.4)$ & $0.6742 \pm 0.0046$ & $0.2329 \pm 0.0004$ \\
$(0.7,\,0.5)$ & $0.5220 \pm 0.0038$ & $0.3118 \pm 0.0008$ \\
$(0.6,\,0.5)$ & $0.5307 \pm 0.0065$ & $0.3092 \pm 0.0004$ \\
$(0.5,\,0.5)$ & $0.5972 \pm 0.0053$ & $0.2750 \pm 0.0004$ \\
\hline
\end{tabular}
\end{table}

In the beginning of the simulation, the center of $D_{1}$ is placed at $(x = 20, y = 50, z = 18)$ and the center of $D_{2}$ is placed at $(x = 68, y = 50, z = 18)$. The droplets are again equilibrated while keeping their mutual interaction off. Once, the system reached equilibrium, $\varepsilon_{d_{1}d_{2}}$ is set to $1.0$ and both droplets start moving in the positive $x-$ direction. Since $\varepsilon_{d_{1}s} > \varepsilon_{d_{2}s}$, $D_{1}$ (trailing droplet) moves faster than the leading droplet $D_{2}$. As a result, on the way both droplets come close enough to come in to physical contact with each other. In this case, the bridge starts forming at the substrate and it then grows in the vertical direction. Since the droplets are fully miscible, the $D_{1}-D_{2}$ interface moves inside $D_{2}$ as the mixing progresses and the end product is a homogeneous mixture of $D_{1}$ and $D_{2}$  particles. Throughout the mixing, the the two droplets continue to move as one entity along the gradient.

The temporal evolution of the bridge height is shown in Fig~\ref{fig:Fig4}. In this case, the values of the power-law  exponents  are $\alpha_{1} = 0.601 \pm 0.016$, and $\alpha_{2} =  0.270 \pm 0.001$ (Figure~\ref{fig:Fig4}). These values are much closer to the power-law exponent reported in the case of stationary coalescence at $K = 20$, but they are still substantially higher than those at $K = 120$ and $K = 240$, the physical location of the coalescence corresponds to the values of $K$ closer to $240$ though. This increase in $\alpha_{1}$ and $\alpha_{2}$ may be attributed to the momentum (inertia) that $D_{1}$ brings with it. In order to understand the coalescence of moving droplets better, we varied the relative
speeds of $D_{1}$ and $D_{2}$. First, the value of $\varepsilon_{d_{1}s}$ was reduced keeping $\varepsilon_{d_{2}s} = 0.40$. Decreasing $\varepsilon_{d_{1}s}$ has two effects, that is the speed of $D_{1}$ reduces, and at the same time, the $D_{1}$-surface attraction also becomes lower. As we observed in the stationary coalescence, $\alpha_{1}$ increases with decreasing droplet--surface attraction, whereas the value of $\alpha_{2}$ decreases. The droplet--surface attraction in the first regime acts against the capillary force thus driving the bridge 
formation, whereas in the second regime it acts against the viscoelastic resistance thus helping in the bridge formation. This trend is also seen here: when the velocity of $D_{1}$ $(\varepsilon_{d_{1}s})$ decreases while keeping $\varepsilon_{d_{2}s}$ fixed, $\alpha_{1}$
increases and $\alpha_{2}$ decreases.  In the case of $\varepsilon_{d_{1}s} = \varepsilon_{d_{2}s} = 0.4$, \textit{e.g.}, the coalescence starts at the equator of the droplet, the value of $\alpha_{1}$ is largest reaches its maximum in this case  while $\alpha_{2}$  acquires its lowest value.

When the attraction of $D_{2}$ with the surface increased 
 to $\varepsilon_{d_{2}s} = 0.50$ while keeping $\varepsilon_{d_{1}s} = 0.70$, the value of $\alpha_{1}$ decreases, whereas the value of $\alpha_{2}$ increases (Table~\ref{tab:table1}). This follwos exactly the opposite trend to the previous case, when velocity of $D_{1}$ decreased. 
The slowdown in the first regime might be attributed to the reduced relative velocity with which $D_{1}$ impacts $D_{2}$ at the contact, thus pushing the bridge inside the $D_{2}$. In the second regime, the increased $D_{2}-$surface attraction drives a faster bridge growth. The bridge height was found to be maximum for the case of $(\varepsilon_{d_{1}s} = \varepsilon_{d_{2}s} = 0.40)$, because both droplets maintained a more spherical shape in this case in comparison with the other cases.

\subsection{Effect of varying \texorpdfstring{$\bm{\varepsilon_{d_1d_2}}$}{epsilon d1 d2}}{\label{resultc}}

\begin{figure}[htbp!]
    \centering
    \setlength{\tabcolsep}{2pt}

    \begin{tabular}{@{}ccccc@{}}
    \includegraphics[width=0.195\textwidth]{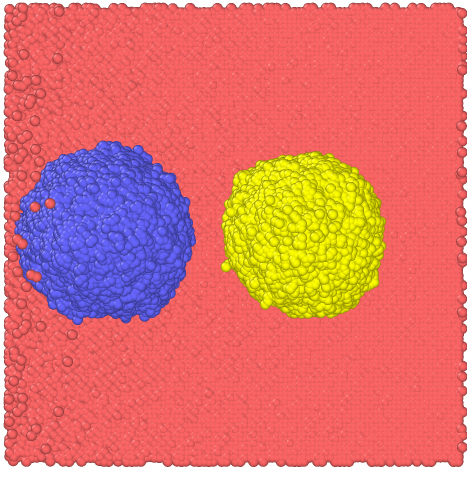} & 
    \includegraphics[width=0.195\textwidth]{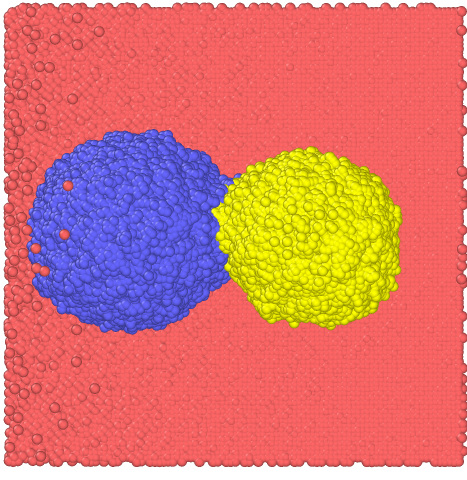} & 
    \includegraphics[width=0.195\textwidth]{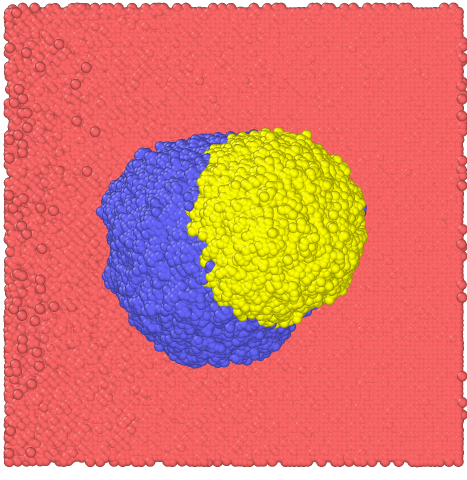} & 
    \includegraphics[width=0.195\textwidth]{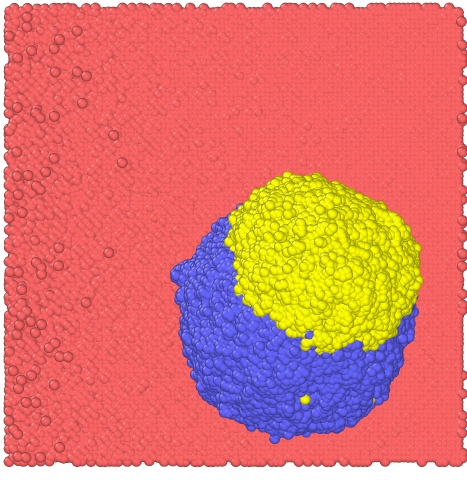} & 
    \includegraphics[width=0.195\textwidth]{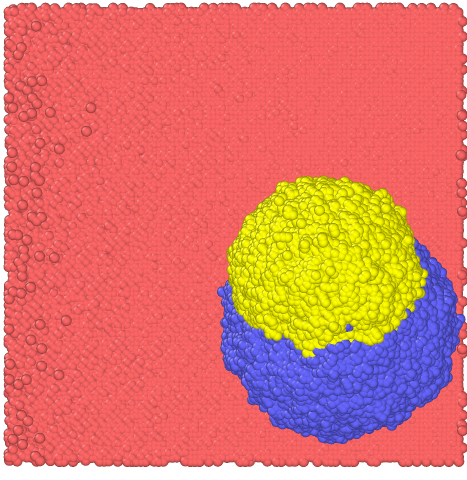} \\
    $t=1.2\times10^5\,\mathrm{ts}$ &
    $t=5.66\times10^5\,\mathrm{ts}$ &
    $t=2.87\times10^6\,\mathrm{ts}$ &
    $t=2.29\times10^7\,\mathrm{ts}$ &
    $t=3.85\times10^7\,\mathrm{ts}$
    \end{tabular}

    \caption{Time-sequenced images of coalescence of two moving droplets for $\varepsilon_{d_1d_2}=0.85$. As can be seen, the trailing droplet tries to bypass the leading one, and the combined entity rotates anti-clockwise. The torque is generated because the the surface$-D_{1}$ attraction is stronger than $D_{1} - D_{2}$ attraction.
    Sec.~\ref{resultc}}
    \label{fig:d12_085}
\end{figure}
\begin{figure}[bp!]
    \centering

    \begin{subfigure}{0.495\textwidth}
        \centering
        \includegraphics[width=\linewidth]{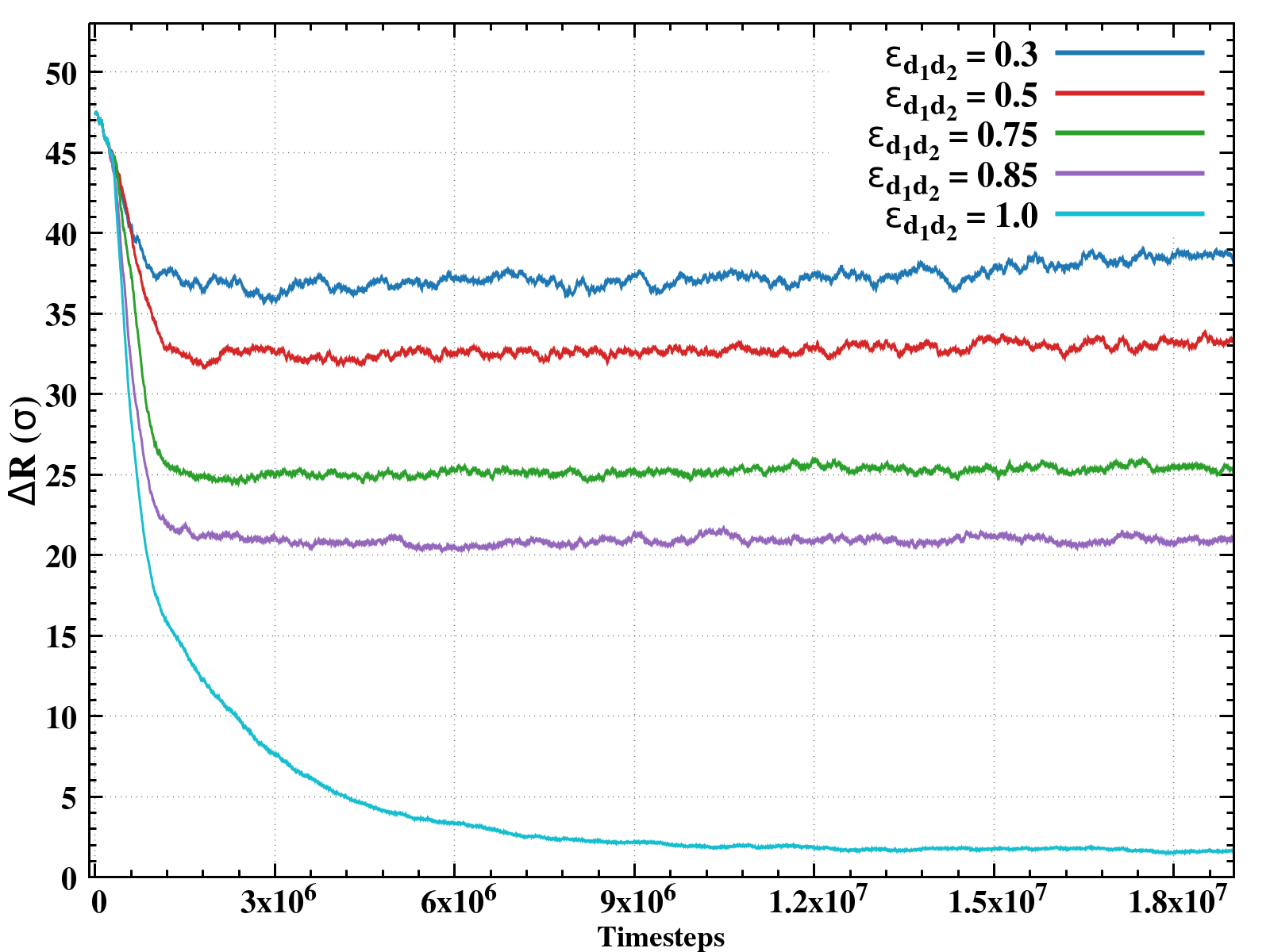}
        \caption{}
        \label{fig:d12a}
    \end{subfigure}
    \begin{subfigure}{0.495\textwidth}
        \centering
        \includegraphics[width=\linewidth]{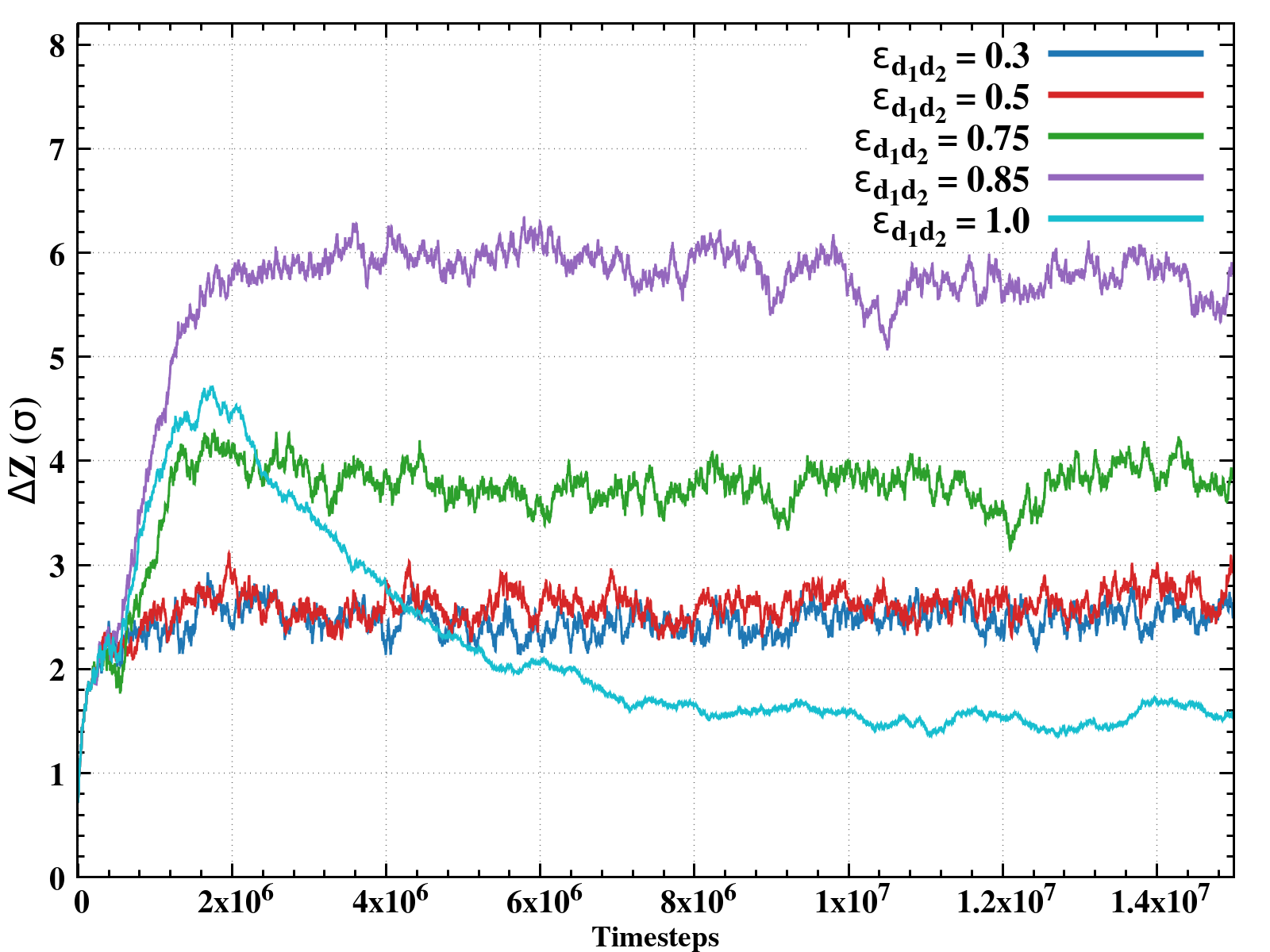}
        \caption{}
        \label{fig:d12b}
    \end{subfigure}

    \caption{Difference of centers of mass of the coalescing droplets ($\Delta R$) and $z-$ component thereof ($\Delta z$) for different $\varepsilon_{d_1d_2}$, other system parameters are the same as  mentioned in Sec.~\ref{resultc}}
    \label{d12}
\end{figure}

In this section, we focus our attention on the coalescence of moving droplets as a function of $\varepsilon_{d_{1}d_{2}}$, which has been varied in the range $1.0 - 0.30$. Other parameters were set to $\varepsilon_{d_1} = \varepsilon_{d_{2}} = 1.0$, $\varepsilon_{d_{1}s} = 0.70$ and $\varepsilon_{d_{2}s} = 0.40$. For $\varepsilon_{d_{1}d_{2}} = 1.0$, the droplets mix completely as we have seen in the previous section. In Fig~\ref{fig:d12_085}, we present the time-sequenced snapshots for the coalescence for $\varepsilon_{d_{1}d_{2}} = 0.85$. In this case, the coalescence does not proceed via a bridge formation. Instead, the $D_{1}$ prefers to continue along the stiffness gradient as $\varepsilon_{d_{1}s} > \varepsilon_{d_{1}d_{2}}$. $D_{2}$, being significantly slower, rather acts like an obstacle, which $D_{1}$ tries to bypass either from the left or from the right. The part of $D_{1}$ that is closer to $D_{2}$ starts to mix with $D_{2}$ with a layer of $D_{1}$ forming between $D_{2}$ and the surface. The combined effect results in a final product that looks like $D_{2}$
partly scooped in  $D_{1}$(Figure~\ref{fig:d12_085}).

\begin{figure}[h]
    \centering
    \includegraphics[width=0.52\linewidth]{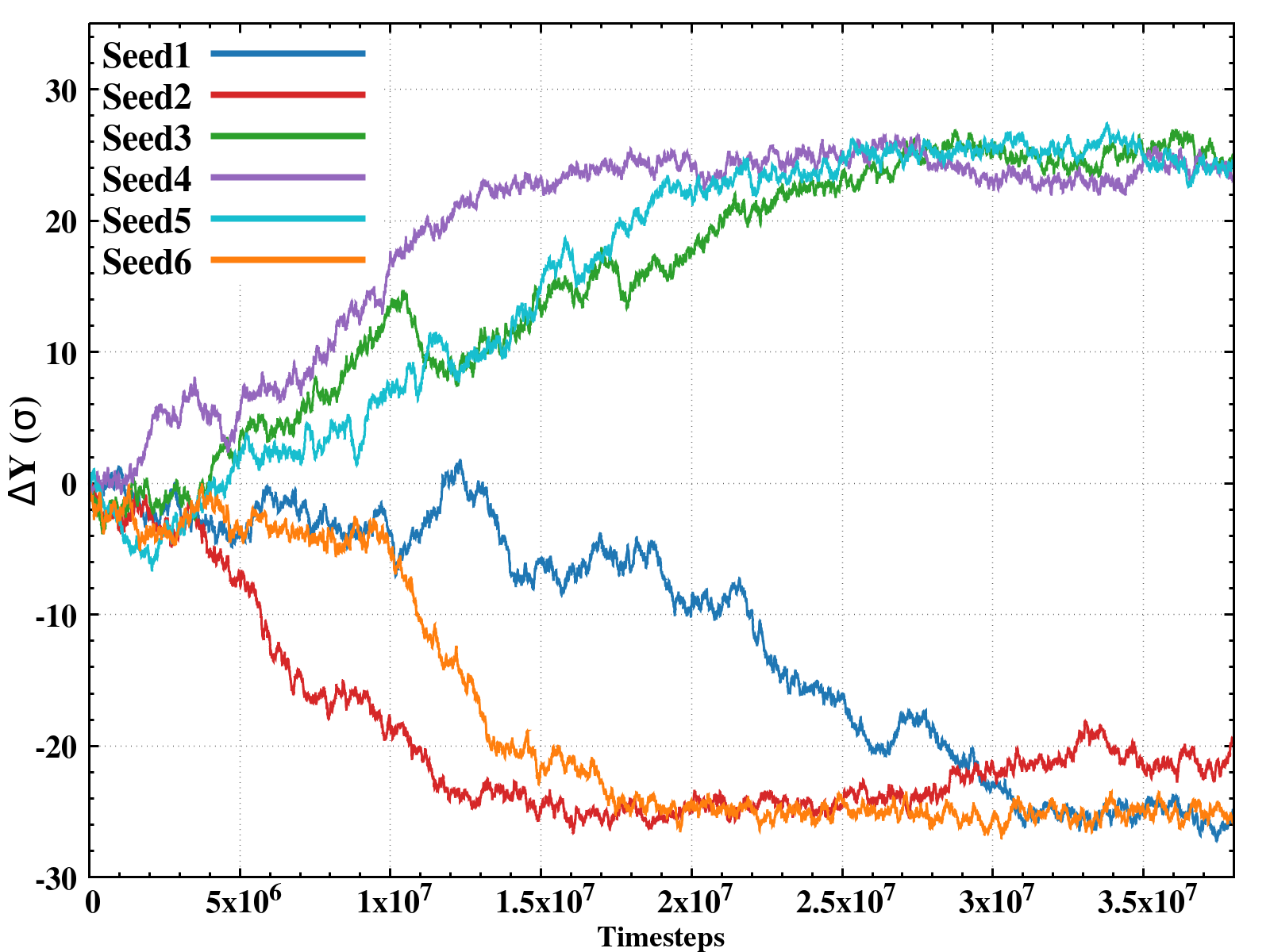}
    \caption{$y-$component of $\Delta R$ for different independent runs, showing that the probability of $D_{1}$ trying to pass $D_{2}$ from left and right are equal. The rest is the same as in Fig.~\ref{fig:d12a}.}
    \label{fig:d12y}
\end{figure}

The value of $\varepsilon_{d_{1}d_{2}}$ was decreased further, the next value considered  was $0.75$. A large $D_{1} - D_{2}$ interface still forms, but there is no layer of $D_{1}$ between 
$D_{2}$ and the surface as the attraction between $D_{1}$ and $D_{2}$ decreases.
The combined body still feels a torque as the part of $D_{1}$ that does not make interface with $D_{2}$ moves faster and tries to overtake $D_{2}$ from either left or right. With decreasing $\varepsilon_{d_{1}d_{2}}$, however, the $D_{1}-D_{2}$ contact area reduces and so does the applied torque. In Fig.~\ref{fig:d12a}, we show the temporal evolution of $\Delta R$ (distance between the centers of mass of $D_{1}$ and $D_{1}$) for varying values of $\varepsilon_{d_{1}d_{2}}$. The distance between the two centers of mass increases with decreasing $\varepsilon_{d_{1}d_{2}}$. The $x$ and $y$ components of $\Delta R$ also increase with decreasing $\varepsilon_{d_{1}d_{2}}$, but the $z$ component varies differently: $\Delta z$ is the maximum for $\varepsilon_{d_{1}d_{2}} = 0.85$ and then decreases when $\varepsilon_{d_{1}d_{2}}$ decreases further (Fig.\ref{fig:d12b}). For $\varepsilon_{d_{1}d_{2}} = 0.85$, a part of $D_{1}$ slides in between the $D_{2}$ and the substrate, but, when $\varepsilon_{d_{1}d_{2}}$ is decreased further, $D_{1}-D_{2}$ interface starts shrinking and the $z-$coordinates of the centers of mass of the droplets align more and more. 
The possibility of $D_{1}$ trying to pass $D_{2}$ from the right is equal to that of $D_{1}$ passing from the left, as there is no force favoring one direction over the other. The same can be seen in the difference between $y-$ coordinates of the centers of mass of $D_{1}$ and $D_{2}$ $(\Delta Y)$ for different independent runs, as shown in Fig~\ref{fig:d12y}.

\subsection{Effect of varying \texorpdfstring{$\bm{\varepsilon_{d_{2}}}$}{epsilon	extsubscript{d2}}}\label{resultd}
\begin{figure}[ht]
    \centering
    \includegraphics[width=0.6\linewidth]{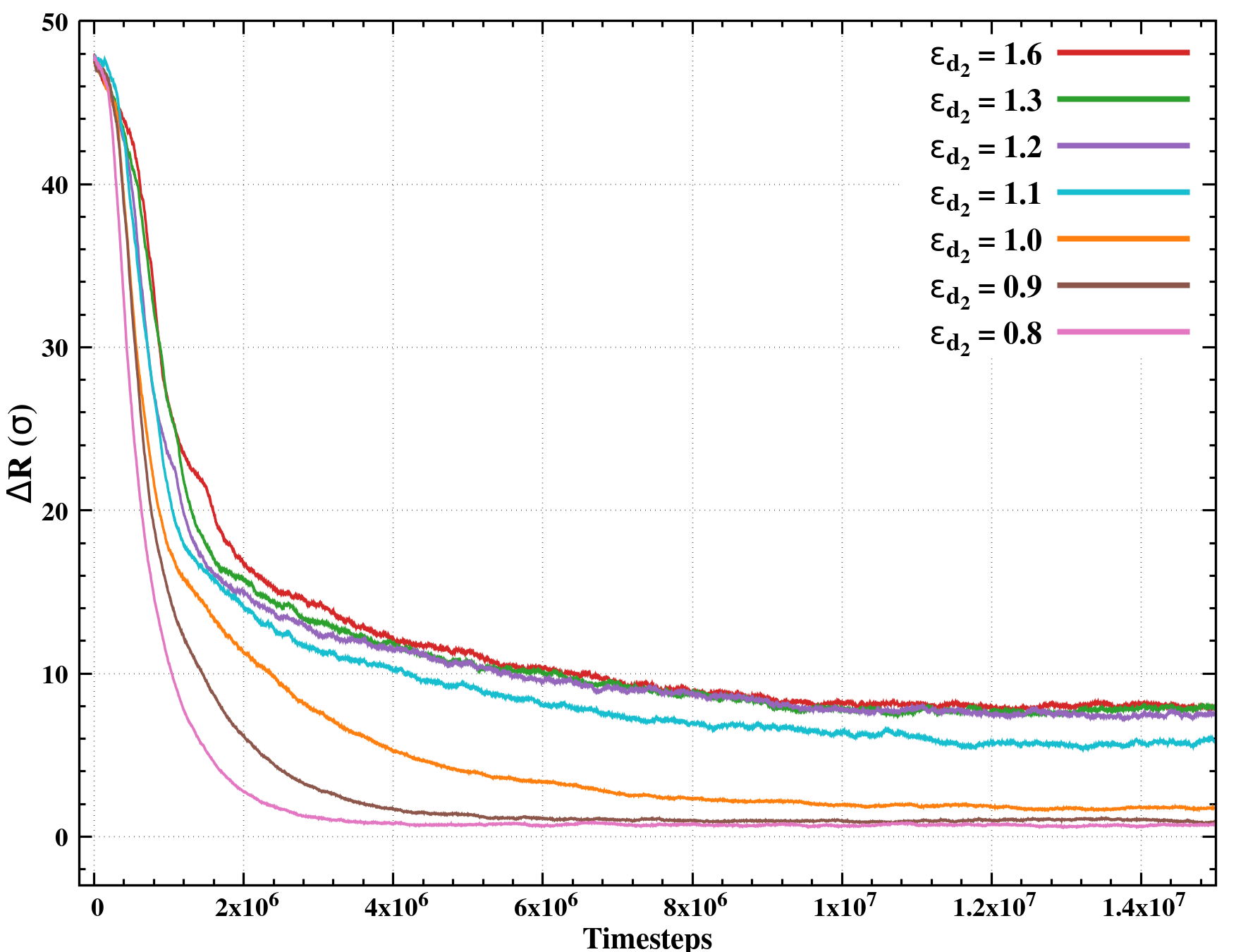}
    \caption{Distance between centers of mass of two coalescing droplets on a gradient surface for varying $\varepsilon_{d_{2}d_{2}}$. As can be seen the distance between the two centers of mass increases with increasing $\varepsilon_{d_{2}d_{2}}$ indicating a decreasing mixing. Rest of the parameters are same as mentioned in the Sec.~\ref{resultd}}
    \label{fig:d2}
\end{figure}
\begin{figure}[htbp!]
    \centering
    \setlength{\tabcolsep}{2pt}

    \begin{tabular}{@{}ccccc@{}}
    \includegraphics[width=0.195\textwidth]{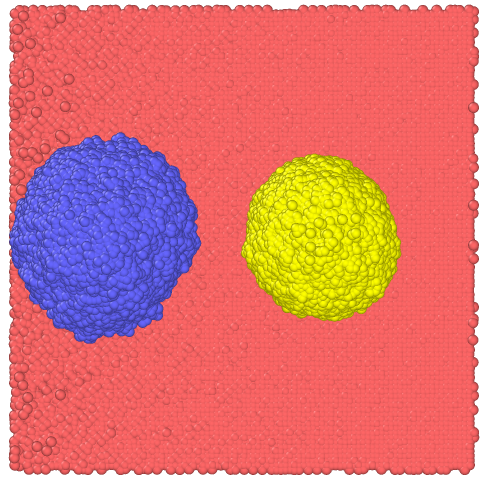} & 
    \includegraphics[width=0.195\textwidth]{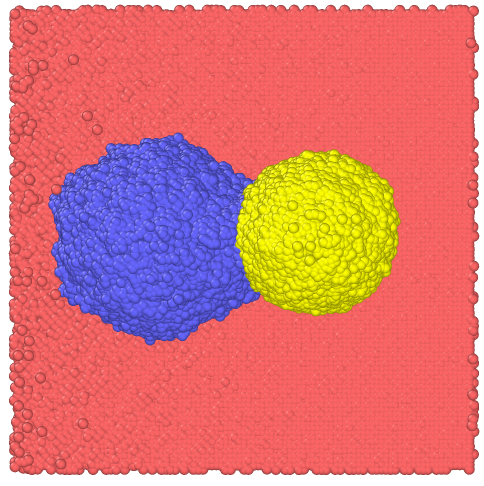} & 
    \includegraphics[width=0.195\textwidth]{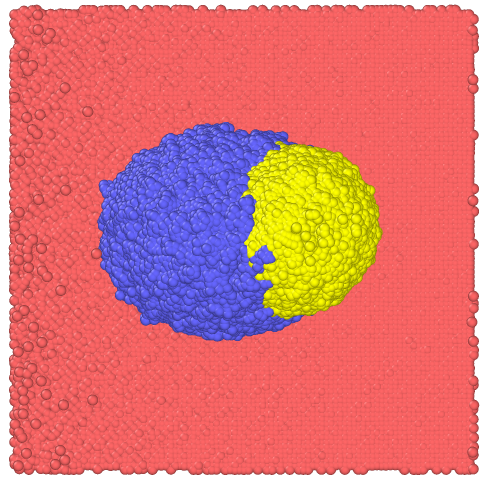} & 
    \includegraphics[width=0.195\textwidth]{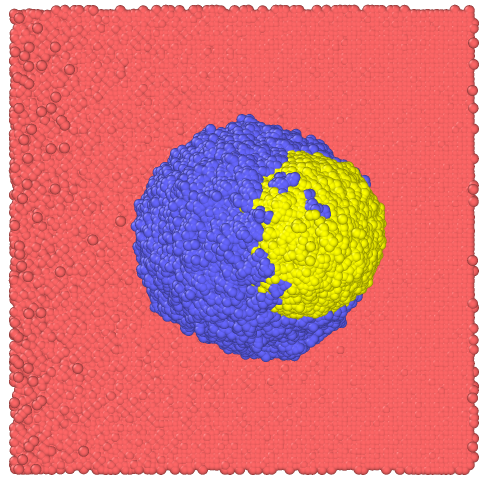} & 
    \includegraphics[width=0.195\textwidth]{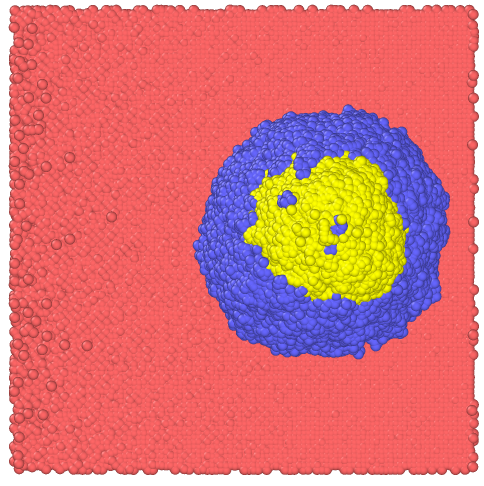} \\
    $t=1.68\times10^5\,\mathrm{ts}$ &
    $t=8.62\times10^5\,\mathrm{ts}$ &
    $t=1.114\times10^6\,\mathrm{ts}$ &
    $t=1.796\times10^6\,\mathrm{ts}$ &
    $t=1.16\times10^7\,\mathrm{ts}$
    \end{tabular}

    \caption{Time sequenced snapshot for coalescence of moving droplets  for $\varepsilon_{d_2}=1.6$. The other parameters are given in Sec. \ref{resultd}}
    \label{fig:d2_1.6}
\end{figure}

In this section, the strength of attraction between the particles comprising $D_{2}$$(\varepsilon_{d_{2}d_{2}})$ was varied keeping $\varepsilon_{d_{1}d_{1}} = \varepsilon_{d_{1}d_{2}} = 1.0$, $\varepsilon_{d_{1}s} = 0.70$, and $\varepsilon_{d_{2}s} = 0.40$. When $\varepsilon_{d_{2}d_{2}}$ is decreased from $1.0$ to $0.90$ and $0.80$, the coalescence becomes faster and the droplets mix to a higher degree. This can be seen in terms of the distance between the centers of mass of the two droplets ($\Delta R$, Fig.~\ref{fig:d2}). The droplets also move along the stiffness gradient while coalescing. When $\varepsilon_{d_{2}d_{2}}$ was increased beyond $1.0$, the coalescence is impeded. Now, the attraction between $D_{2}$ particles is stronger than that between the particles of $D_{1}$, as well as between the particles of $D_{1}$ and $D_{2}$, so the mixing is no more uniform. The particles belonging to $D_{2}$ droplet try to stay together. As a result, the droplet moves in a way that it keeps its structure inside $D_{1}$  as illustrated by the partial mixing inside the envelope of $D_{1}$. With increasing $\varepsilon_{d_{2}d_{2}}$, the depth of the $D_{1}$ envelope on $D_{2}$ becomes smaller as the degree of mixing decreases, due to the increased attraction between the $D_{2}$ particles. This, again, is reflected in $\Delta R$, which increases with increasing $\varepsilon_{d_{2}d_{2}}$ (Fig.~\ref{fig:d2}). 
In Fig.~\ref{fig:d2_1.6}, the time-sequenced snapshots for $\varepsilon_{d_{1}d_{2}} = 1.6$ are presented to provide a better understanding of  the final product when the inter-droplet attraction of one droplet is much stronger than the inter-droplet attraction of the other droplet.

\section{Summary and Future Outlook}{\label{summary}}
In this manuscript, we have studied the coalescence of polymeric droplets on a soft surface, by means of molecular dynamics simulation of a coarse-grained model. First, we studied the coalescence of two stationary droplets on a soft surface without stiffness gradient for three different values of ``surface-softness''.
The coalescence dynamics were characterized by analyzing the temporal growth of the bridge height, which unlike Newtonian fluids, exhibits two distinct regimes of power-law dependence. A faster bridge growth, in the beginning, is followed by a slower growth before attaining a constant value. The slowdown of the dynamics in the second regime might be attributed to the emergence of viscoelastic relaxation of the polymer chains. The bridge height grows the fastest in the first regime for the softest surface and  slows down with increasing stiffness. In the second regime,
the trend is reversed, {\it i.e,} the rate of the bridge growth increases with increasing stiffness. This may be understood in terms of the competition between the droplet--droplet attraction, the droplet--surface attraction and the viscoelastic relaxation. Then, a stiffness gradient was created on the surface and the droplets were initially kept sufficiently apart. The droplets undergo durotactic motion in response to the gradient, moving from the softer 
to the stiffer end of the substrate. The velocity of the droplets depends on the droplet--surface attraction: $\varepsilon_{d_{1}s}$ and $\varepsilon_{d_{2}s}$ are chosen such that the trailing droplet moves with a higher velocity in comparison with the leading droplet. On the way, when the droplets come close enough to interact, coalescence starts  with the combined entity contuning the durotactic motion. The dynamics and the extent of the coalescence depend on a number parameters, which will be summarized in the following. 

The speed of coalescing droplets was varied and it was found that decreasing the velocity of the trailing droplet favors the bridge growth in the first regime, whereas it slows it down in the second regime. The opposite effect was observed  when increasing the velocity of $D_{2}$, {\it i.e, }
the bridge height growth becomes slower in the first regime and faster in the second regime. This may, again, be understood in terms of the competition between capillary forces, viscoelastic resistance, and relative momentum of the moving droplets. A slower $D_{1}$ in the beginning means smaller relative momentum, but, also, a weaker $D_{1} -$ surface attraction to compete with the $D_{1} - D_{2}$ attraction. In contrast, in the second regime, a weaker $D_{1} -$ surface attraction  reflects a weaker competition to the chain relaxation. Similarly, a faster $D_{2}$ or a higher 
value of $\varepsilon_{d_{2}s}$ means the particles forming bridge will feel more attracted to the surface and hence the bridge growth slows down, in the first regime. In the second regime, higher 
$\varepsilon_{d_{2}s}$ will oppose viscoelastic relaxation, leading to a faster growth.

In addition, the strength of the inter-droplet interaction was varied.
When  $\varepsilon_{d_{1}d_{2}}$ was increased beyond $\varepsilon_{d_{1}d_{1}} = \varepsilon_{d_{2}d_{2}} = 1.0$, the coalescence becomes faster and the droplets mix better. However, when $\varepsilon_{d_{1}d_{2}}$ was decreased below $\varepsilon_{d_{1}} = \varepsilon_{d_{2}d_{2}} = 1.0$, the droplets mix up to a certain degree. In particular, the degree of mixing decreases with decreasing $\varepsilon_{d_{1}d_{2}}$. The combined body of the two droplets also experiences torque while coalescing and moving at the same time. So far, the value of $\varepsilon_{d_{1}d_{1}}$ was kept 
equal to $\varepsilon_{d_{2}d_{2}}$. When $\varepsilon_{d_{2}d_{2}}$ was taken to be greater than $\varepsilon_{d_{1}d_{1}}$ and $\varepsilon_{d_{1}d_{2}}$, the coalescence  exhibits a slightly different behavior. The $D_{1}$ droplet, while coalescing with $D_{2}$, also envelops $D_{2}$. Since the attraction between $D_{2}$ particles is stronger than that between $D_{1}$ particles and the attraction between $D_{1} - D_{2}$ particles, the $D_{2}$ opposes mixing. The envelope is created by the $D_{1}$ particles, which  are pulled  into $D_{2}$, due to capillary forces. Both the extent of mixing and enveloping decrease with increasing $\varepsilon_{d_{2}d_{2}}$. 

Apart from engineering applications, the findings of this study may carry relevance for our fundamental understanding of biological processes, particularly in the context of fusion of multicellular aggregates, which plays an important role in development, disease, and therapy. The demonstration that durotaxis-driven coalescence of viscoelastic droplets produces qualitatively distinct morphological outcomes --- complete mixing, partial enveloping, or bypass --- depending on the ratio of the inter- and intra-droplet cohesion maps naturally onto the collective behavior of cell clusters whose intercellular adhesion strength  differs between cluster types. The enveloping morphology, in particular, draws a parallel with the engulfment of mechanically stiffer tumor spheroids by softer stromal or immune cell aggregates. It is also worth mentioning here that fusion of cellular aggregates during morphogenesis the fusion exhibits two regimes: an initial rapid contact followed by a much slower, relaxation-limited integration, which is sensitive to the mechanical properties of the substrate.

We anticipate that the results of this study may offer  new insights into the coalescence of viscoelastic droplets on soft substrates. However, there is still a lot to be explored when it comes to the coalescence of 
non-Newtonian fluids, coalescence on soft surfaces, or coalescence of moving droplets. It would be interesting, for example, to study the effect of bending rigidity of the droplets on the coalescence. Another aspect, that is relevant for microfluidics as well as in a biological context is to study the coalescence of viscoelastic fluids in  confined geometries, such as micro-channels or a Hele-Shaw geometry. 

\section*{Acknowledgment}
DT, VK and SLS acknowledge the generous financial support received from BHU under the IoE scheme
and from SERB, New Delhi. DT and SLS also thank CSIR, New Delhi for financial support
(grant id: 03WS-012-2023-24-EMR-II).
\bibliographystyle{apsrev4-2}  
\bibliography{reference}
\end{document}